\documentclass[12pt]{article}
\begin{document}
\title{{\bf{\ Corrected entropy of BTZ black hole in tunneling approach}}}
\author{
  {\bf {\normalsize Sujoy Kumar Modak}$
 $\thanks{E-mail: sujoy@bose.res.in}}\\
 {\normalsize S.~N.~Bose National Centre for Basic Sciences,}
\\{\normalsize JD Block, Sector III, Salt Lake, Kolkata-700098, India}
\\[0.3cm]
}
\date{}

\maketitle

{\bf Abstract:}\\ 
We investigate further the recent analysis \cite{R.Banerjee2}, based on a Hamilton-Jacobi type approach, to compute the temperature and entropy of black holes beyond the semiclassical approximation. It is shown how non spherically symmetric geometries are inducted in the general formalism by explicitly considering the BTZ black hole. The leading (logarithmic) and non leading corrections to the area law are obtained. \\

The fact that black holes can radiate \cite{Hawking,Hawking1} has led to a calculation of its thermodynamic entities like temperature and entropy. The celebrated Bekenstein-Hawking area law governing the entropy ($S$) of a black hole in terms of its horizon area ($A$), $$S=\frac{A}{4}$$ has been derived, in the semiclassical approximation, following various approaches \cite{Hawking1,Beken,Beken1,Hawking2,Beken2,Hawking3}.

The area law undergoes corrections due to quantum effects. These corrections have been computed using field theory methods \cite{Fursaev, Mann, Page}, quantum geometry techniques \cite{Partha,Ben}, general statistical mechanical arguments \cite{Partha1,More,Pal}, Cardy formula \cite{Carlip,Setare} etc. It appears that the leading correction is logarithmic while nonleading terms involve inverse powers of area $A$.

Now a particularly illuminating way of discussing the Hawking effect, which also fits in with the general intuitive picture, is to use the tunneling formalism \cite{Wilczek,Cai,Padma,Kerner,Kerner1}. The semiclassical Hawking temperature has been computed following either the null geodesic \cite{Wilczek,Cai} or Hamilton-Jacobi \cite{Padma,Kerner,Kerner1} variants of this formalism. Among the various methods of obtaining the corrections to the semiclassical expressions mentioned in the preceding paragraph, a systematic analysis within the tunneling approach was missing. This has been recently discussed in a series of papers by Banerjee $\it{et}$ $ \it{ al}$ \cite{R.Banerjee1,R.Banerjee2,R.Banerjee3,R.Banerjee4}. However, explicit examples were mostly confined to black holes with spherically symmetric geometries.

The object of this paper is to investigate further the general formalism of \cite{R.Banerjee2} by including non spherically symmetric geometries. We explicitly compute the corrected expressions for the temperature, entropy and area law for the BTZ black hole. Apart from the standard semiclassical structures, the leading (logarithmic) as well as non leading corrections to the area law are also reproduced. While the leading correction has been obtained earlier using other methods \cite{Carlip,Govindarajan}, the non leading corrections have not been discussed.

The metric for the (2+1) dimensional BTZ black hole with negative cosmological constant $\Lambda= -\frac{1}{l^2}$ and in the unit of $8G_3=1$, where $G_3$ is the three dimensional Newton's constant, is given by \cite{Banados},   
\begin{eqnarray}
ds^2 = -N^2dt^2+N^{-2}dr^2+r^2(N^\phi dt+d\phi)^2,
\label{1.01}
\end{eqnarray}
with the lapse function
\begin{eqnarray}
N^2(r) = -M+\frac{r^2}{l^2}+\frac{J^2}{4r^2},
\label{1.02}
\end{eqnarray}
and
\begin{eqnarray} 
N^\phi (r) = -\frac{J}{2r^2}. 
\label{1.03}
\end{eqnarray}
M and J are respectively the mass and angular momentum of the BTZ black hole. Outer (event) and inner horizons are obtained by setting $g^{rr}=N^2= 0$. This yields, 
\begin{eqnarray} 
r_{\pm} = \frac{l}{\sqrt 2}\left(M\pm\sqrt{M^2-\frac{J^2}{l^2}}\right)^{1/2}.
\label{1.04}
\end{eqnarray}
Therefore the horizon radius is a function of both M and J. The area of the event (outer) horizon \cite{Carlip1,Carlip2} is given by, 
\begin{equation}
A=2 \pi r_+.
\label{1.041}
\end{equation}
The angular velocity for BTZ black hole follows from the general expression for angular velocity for any rotating black hole, given by \cite{Carrol},
\begin{eqnarray}
\Omega= \Big[-\frac{g_{\phi t}}{g_{\phi \phi}}-\sqrt{{(\frac{g_{t\phi}}{g_{\phi\phi}})^2}-\frac{g_{tt}}{g_{\phi \phi}}}\Big]_{r=r_+}.
\label{rev1}
\end{eqnarray}
For BTZ black hole (\ref{rev1}) reduces to,
\begin{eqnarray}
\Omega= -\frac{g_{\phi t}}{g_{\phi \phi}}\Big|_{r=r_+}=\frac{J}{2r^2_+}.
\label{1.06}
\end{eqnarray}

The first step in the analysis is to compute the semiclassical Hawking temperature. For this purpose we adopt the tunnelling formalism \cite{Wilczek,Cai,Padma,Kerner,Kerner1,Ang,R.Banerjee2} within the Hamilton-Jacobi approach. It is thus desirable to isolate the $r-t$ sector of the metric (\ref{1.01}). Therefore we first make the following transformation,

\begin{eqnarray} 
d\chi=d\phi-\Omega dt
\label{1.05}
\end{eqnarray}
i.e, 
\begin{eqnarray}
\chi=\phi-\Omega t.
\label{new11}
\end{eqnarray}
Such a transformation was used earlier in \cite{Kerner1,Ang}. In the near horizon approximation, using the transformation (\ref{new11}) the metric(\ref{1.01}) can be written in the desired form,
\begin{eqnarray}
ds^2 = -N^2dt^2+N^{-2}dr^2+r_+^2d\chi^2,
\label{new1}
\end{eqnarray}
 where the $r-t$ sector is isolated from the angular part ($d\chi^2$).

A general procedure based on the Hamilton-Jacobi method for calculation of Hawking temperature is done in \cite{R.Banerjee2}. In this method one needs to take only the radial trajectory so that the $r-t$ sector of the metric as written in the form (\ref{new1}) is relevant. The expression of semiclassical Hawking temperature for any general metric is now given by \cite{R.Banerjee2}, 
\begin{eqnarray}
T=\frac{\hbar}{4}{\left(Im\int_C{\frac{dr}{\sqrt {g_{tt}(r)g^{rr}(r)}}}\right)}^{-1}
\label{new3}
\end{eqnarray} 
Considering the metric (\ref{new1}), the Hawking temperature for the BTZ black hole is found to be,
\begin{eqnarray}
T=\frac{\hbar}{4}{\left(Im\int_C{\frac{dr}{N^2(r)}}\right)}^{-1}=\frac{\hbar}{4}{\left(l^2Im\int_C{\frac{r^2dr}{(r^2-r^2_+)(r^2-r^2_-)}}\right)}^{-1}.      
\label{new4}
\end{eqnarray}
The integration is to be performed remembering that the particle tunnels from just behind the event horizon to the outer region. So the integrand has a simple pole at $r=r_+.$ Choosing the contour as a half loop going around this pole from left to right and integrating, we obtain the desired result for the semiclassical Hawking temperature, 
\begin{eqnarray}
T=\frac{\hbar}{2\pi l^2}\left(\frac{r^2_+-r^2_-}{r_+}\right).
\label{1.07}
\end{eqnarray}
This is the semiclassical Hawking temperature and agrees with the results quoted in literature \cite{Carlip1,Carlip2}. Substituting the values of $r_+$ and $r_-$ from (\ref{1.04}) we get,
\begin{eqnarray} 
T =\frac{\sqrt 2\hbar}{2\pi l}\frac{{\sqrt{M^2-\frac{J^2}{l^2}}}}{\left(M+\sqrt{M^2-\frac{J^2}{l^2}}\right)^{1/2}}.
\label{1.071}
\end{eqnarray}
 The semiclassical Hawking temperature can also be derived by using the near horizon expansion of metric coefficients as,$$g_{tt}(r)=g_{tt}(r_+)+(r-r_+)(g_{tt})'(r_+)=(r-r_+)(g_{tt})'(r_+)$$ and $$g^{rr}(r)=g^{rr}(r_+)+(r-r_+)(g^{rr})'(r_+)=(r-r_+)(g^{rr})'(r_+).$$ Using this expansion in (\ref{new3}) it follows that,
\begin{eqnarray}
T={\frac{\hbar\sqrt{g'_{tt}(r_+)g'^{rr}(r_+)}}{4\pi}}.
\label{}
\end{eqnarray}
 This expression of semiclassical Hawking temperature was reported earlier in \cite{Kerner1,Ang}.

After finding the semiclassical Hawking temperature we proceed with the calculation of the semiclassical Bekenstein-Hawking entropy. The corrections to these semiclassical results will be considered later. Consider the following law of thermodynamics for a chargeless rotating black hole,
\begin{eqnarray}
dM= TdS+\Omega dJ,
\label{1.08}
\end{eqnarray}
which can be written in the form
\begin{eqnarray}
dS(M,J)= \frac{dM}{T}-\frac{\Omega}{T}dJ.
\label{1.09}
\end{eqnarray}
Inverting the Hawking temperature (\ref{1.071}) of the BTZ black hole, we obtain,
\begin{eqnarray}
\frac{1}{T}(M,J)= \frac{\sqrt 2\pi l \left(M+\sqrt{M^2-\frac{J^2}{l^2}}\right)^{1/2}}{\hbar\sqrt{M^2-\frac{J^2}{l^2}}}.
\label{1.10}
\end{eqnarray} 
The angular velocity (\ref{1.06}) simplifies to,
\begin{eqnarray}
\Omega=-\frac{g_{\phi t}}{g_{\phi\phi}}|_{r=r_+}=\frac{J}{l^2(M+\sqrt{M^2-\frac{J^2}{l^2}})}.
\label{1.101}
\end{eqnarray}
Using (\ref{1.10}) and (\ref{1.101}) yields,
\begin{eqnarray}
\frac{\Omega}{T}(M,J)= \frac{\sqrt 2 \pi J}{l\hbar\left(M+\sqrt{M^2-\frac{J^2}{l^2}}\right)^{1/2}\sqrt{M^2-\frac{J^2}{l^2}}}.
\label{1.11}
\end{eqnarray}
We identify (\ref{1.09}) as a first order differential equation. Further we note that $dS$ in (\ref{1.09}) is an exact differential since it satisfies the relation,
\begin{eqnarray}
\frac{\partial}{\partial J}\left(\frac{1}{T}\right)=\frac{\partial}{\partial M}\left(\frac{-\Omega}{T}\right),
\label{new5}
\end{eqnarray}
 as may be checked on exploiting the equations (\ref{1.10}), (\ref{1.11}). Now, in order to calculate the entropy ($S$), (\ref{1.09}) has to be solved. Here we follow the general procedure for solving any exact first order differential equation, 
\begin{eqnarray}
df(x,y)= P(x,y)dx + Q(x,y)dy.
\label{1.12}
\end{eqnarray}
$df(x,y)$ in (\ref{1.12}) is an exact differential if $P$ and $Q$ satisfy the condition
\begin{eqnarray}
\frac{\partial P}{\partial y}= \frac{\partial Q}{\partial x}.
\label{1.13}
\end{eqnarray}
If (\ref{1.13}) holds then the solution of (\ref{1.12}) is given by
\begin{eqnarray}
f(x,y)= \int Pdx+\int Qdy- \int\frac{\partial}{\partial y}\left(\int Pdx\right)dy.
\label{1.14}
\end{eqnarray}

Comparing (\ref{1.09}) with (\ref{1.12}) we obtain the following correspondence, 
\begin{eqnarray}
x\rightarrow M, y\rightarrow J, P\rightarrow \frac{1}{T}, Q\rightarrow \frac{-\Omega}{T}, f\rightarrow S.
\label{new6}
\end{eqnarray} 
Using this dictionary it is very easy to write the solution of (\ref{1.09}) for the semiclassical result for entropy of the spinning BTZ black hole,
\begin{eqnarray}
S= \int\frac{1}{T}dM + \int\frac{-\Omega}{T}dJ -\int\frac{\partial}{\partial J}\left(\int\frac{1}{T}dM\right)dJ.
\label{1.16}
\end{eqnarray}
The solution of the integral over $dM$ gives
\begin{eqnarray}
\int\frac{1}{T}dM= 4\pi \frac{l}{\sqrt 2 \hbar}(M+\sqrt{M^2-\frac{J^2}{l^2}})^{1/2}=\frac{4\pi r_+}{\hbar}. 
\label{1.17}
\end{eqnarray}
Having this result, it can be easily checked that the following relation holds,
\begin{eqnarray} 
\frac{\partial}{\partial J}\left(\int\frac{dM}{T}\right)=\frac{-\Omega}{T}.
\label{1.18}
\end{eqnarray}
Substituting the above relation in (\ref{1.16}) and using (\ref{1.17}), immediately leads to the entropy of the spinning BTZ black hole,
\begin{eqnarray}
S=\int\frac{1}{T}dM=\frac{4\pi r_+}{\hbar}. 
\label{1.19}
\end{eqnarray}
If we write this in terms of horizon area (\ref{1.041}) by reinstating the unit $8G_3=1$ , we get
\begin{eqnarray}
S=\frac{A}{4\hbar G_3}=S_{{\textrm {BH}}}, 
\label{1.1911}
\end{eqnarray}
 which is the well known semiclassical Bekenstein-Hawking area law \cite{Hawking1,Beken,Hawking2}.

In order to find the entropy beyond the semiclassical approximation, quantum corrections to the Hawking temperature must be included. To compute this we shall follow the analysis in \cite{R.Banerjee2}, developed for any static, spherically symmetric black hole. The BTZ metric as given in (\ref{new1}) is static. In the near horizon approximation and passing to the coordinate $\chi$ one can isolate the $r-t$ sector of the metric (\ref{new1}) from the angular part ($d\chi^2$). Therefore, we can follow the analysis of \cite{R.Banerjee2} for (\ref{new1}). The massless particle in spacetime (\ref{new1}), governed by the Klein-Gordon equation is given by,
\begin{equation}
-\frac{\hbar^2}{\sqrt{-g}}{\partial_\mu[g^{\mu\nu}\sqrt{-g}\partial_{\nu}]\phi}=0.
\label{rev2}
\end{equation}  
 Since we are concerned about the radial trajectory, only $r-t$ sector of the metric (\ref{new1}) is relevant. In this case (\ref{rev2}) takes the form,
\begin{eqnarray}
-\frac{1}{N^2(r)}{\partial_t^2\phi}+N^2(r){\partial_r^2\phi}+\partial_r(N^2(r)){\partial_r\phi}=0.
\label{rev3}
\end{eqnarray}         
The semiclassical wave function ($\phi$) which satisfy the Klein-Gordon equation is given by,
\begin{eqnarray}
\phi(r,t)=exp[-\frac{i}{\hbar}{{\cal S}(r.t)}]. 
\label{rev4}
\end{eqnarray}
Where, ${\cal {S}}(r,t)$ is one particle action and it will be expanded in the powers of $\hbar$. Putting $\phi(r,t)$ from (\ref{rev4}) in (\ref{rev3}) we get,
\begin{equation}
{\frac{1}{N^2}{(\partial_t {\cal {S}}(r,t))^2}-N^2(\partial_r {\cal S}(r,t) )^{2}- \frac{\hbar}{i}{[\frac{1}{N^2}{\partial^2_t {\cal {S}}(r,t)}-N^2{\partial^2_r {\cal {S}}(r,t)}}
-{\partial_r N^2(r)}\partial_r {\cal S}(r,t)]}=0.
\label{rev5}
\end{equation} 
An expansion of ${\cal S}(r,t)$ in powers of $\hbar$ gives,
\begin{eqnarray}
{\cal S}(r,t)={\cal S}{_0}(r,t)+  \sum_i{\hbar^i {\cal S}_i(r,t)},
\label{rev6}
\end{eqnarray}
 where $i=1,2,3...$. The ${\cal O}(\hbar)$ corrections are treated as quantum corrections to the semiclassical value of one particle action ${\cal S}(r,t)$. Substituting (\ref{rev6}) in (\ref{rev5}) and simplifying we find the following set of equations for different powers of $\hbar$,

\begin{eqnarray}
\hbar^0~:~\frac{\partial S_0}{\partial t}=\pm N^2(r)\frac{\partial S_0}{\partial r},
\label{rev7}
\end{eqnarray}
\begin{eqnarray}
\hbar^1~:~&&\frac{\partial S_1}{\partial t}=\pm N^2(r)\frac{\partial S_1}{\partial r},
\nonumber
\\
\hbar^2~:~&&\frac{\partial S_2}{\partial t}=\pm N^2(r)\frac{\partial S_2}{\partial r},
\nonumber
\\
.
\nonumber
\\
.
\nonumber
\\
.
\nonumber
\end{eqnarray} 
and so on. The fact that ${\cal S}_i(r,t)$'s are proportional to ${\cal S}_0(r,t)$ is revealed from the similar functional dependence of these linear differential equations. Considering this, the corrected version of one particle action is given by,
\begin{equation}
{\cal{S}}(r,t)=(1+\sum_i\gamma_i\hbar^i){\cal{S}}_0(r,t),
\label{1.191}
\end{equation} 
 where ${\cal S}_0$ denotes the semiclassical contribution. This shows that $\gamma_i$'s have the dimension $\hbar^{-i}$. To make these proportionality constants dimensionless we consider the following dimensional analysis. In $(2+1)$ dimensions in the units $8G_3=1$ and $c=k_B=1$ the Planck constant ($\hbar$) is of the order of Planck length ($l_P$) \cite{Carlip3}. Since the only length parameter for the black hole is $r_+$, we can write (\ref{1.191}) as
\begin{equation}
{\cal{S}}(r,t)=(1+\sum_i\frac{\beta_i\hbar^i}{r^i_+}){\cal{S}}_0(r,t),
\label{1.192}
\end{equation}
 where $\beta_i$'s are dimensionless constants. Due to the symmetry of the metric (\ref{1.01}) one is looking for the solution for the semiclassical action $({\cal O}(\hbar))$ in the form,
\begin{equation}
{\cal{S}}_0=Et+J\phi+\tilde{{\cal{S}}}_0(r).
\label{rev7a}
\end{equation}
 In the near horizon approximation using (\ref{new11}) and considering the $r-t$ sector of the metric (\ref{new1}), the form of semiclassical action takes the form,
\begin{equation}
{\cal{S}}_0(r,t)=\omega t + \tilde {\cal S}_0(r),
\label{rev8}
\end{equation}
 where $\omega=E+J\Omega$ is considered as the energy of the emitted particle. Substituting this in (\ref{rev7}) and integrating we get,
\begin{eqnarray}
\tilde {\cal S}_0(r)= \pm \omega\int_C\frac{dr}{N^2(r)}.
\label{rev9} 
\end{eqnarray}
 The $+ (-)$ sign indicates that the particle is ingoing (outgoing). Using (\ref{rev9}) and (\ref{rev8}), equation (\ref{1.192}) can be written as,
\begin{equation}
{\cal{S}}(r,t)=(1+\sum_i\frac{\beta_i\hbar^i}{r^i_+}){\cal{S}}_0(r,t)=(1+\sum_i\frac{\beta_i\hbar^i}{r^i_+})(\omega t \pm \omega\int_C\frac{dr}{N^2(r)}).
\label{rev10}
\end{equation}

Therefore solution for the ingoing and outgoing particle of the Klein-Gordon equation under the background metric (\ref{new1}) are respectively,
\begin{eqnarray}
\phi_{{\textrm {in}}}= {\textrm{exp}}\Big[-\frac{i}{\hbar}(1+\sum_i\beta_i\frac{\hbar^i}{r^i_+})\Big(\omega t  +\omega\int_C\frac{dr}{N^2(r)}\Big)\Big]
\label{rev11}
\end{eqnarray} 
\begin{eqnarray}
\phi_{{\textrm {out}}}= {\textrm{exp}}\Big[-\frac{i}{\hbar}(1+\sum_i\beta_i\frac{\hbar^i}{r^i_+})\Big(\omega t  -\omega\int_C\frac{dr}{N^2(r)}\Big)\Big].
\label{rev12}
\end{eqnarray} 
 In the tunneling formalism the ingoing particle crosses the horizon classically and enters in the black hole region whereas the outgoing particle just tunnels through the horizon to the outer region. The path for outgoing particle is classically forbidden. This is precisely because of the fact that the metric coefficients in the $r-t$ sector changes sign for the tunneling of the outgoing particle. The path in which tunneling takes place has imaginary time coordinate ($Im~t$). Therefore, ingoing and outgoing probabilities are given by,  
\begin{eqnarray}
P_{{\textrm{in}}}=|\phi_{{\textrm {in}}}|^2= {\textrm{exp}}\Big[\frac{2}{\hbar}(1+\sum_i\beta_i\frac{\hbar^i}{r^i_+})\Big(\omega{\textrm{Im}}~t +\omega{\textrm{Im}}\int_C\frac{dr}{N^2(r)}\Big)\Big]
\label{rev13}
\end{eqnarray}
and
\begin{eqnarray}
P_{{\textrm{out}}}=|\phi_{{\textrm {out}}}|^2= {\textrm{exp}}\Big[\frac{2}{\hbar}(1+\sum_i\beta_i\frac{\hbar^i}{r^i_+})\Big(\omega{\textrm{Im}}~t -\omega{\textrm{Im}}\int_C\frac{dr}{N^2(r)}\Big)\Big].
\label{rev14}
\end{eqnarray}
 In the classical limit ($\hbar\rightarrow 0)$ the ingoing probability is unity, hence,
\begin{eqnarray}
{\textrm{Im}}~t = -{\textrm{Im}}\int_C\frac{dr}{N^2(r)}.
\label{rev15}
\end{eqnarray}
 Therefore probability for the outgoing particle is,
\begin{eqnarray}
P_{{\textrm{out}}}={\textrm{exp}}\Big[-\frac{4}{\hbar}{\omega}(1+\sum_i\beta_i\frac{\hbar^i}{r^i_+}){\textrm{Im}}\int_C\frac{dr}{N^2(r)}\Big].
\label{rev16}
\end{eqnarray}
Now, the temperature of the BTZ black hole is obtained by using the principle of ``detailed balance'' \cite{Padma}, which states,
\begin{eqnarray}
\frac{P_{{\textrm{out}}}}{P_{\textrm{in}}}= {\textrm {exp}}\Big(-\frac{\omega}{T_h}\Big).
\label{rev17}
\end{eqnarray}
 Taking $P_{\textrm {in}}$ to be unity it follows that,
\begin{eqnarray}
P_{{\textrm{out}}}= {\textrm {exp}}\Big(-\frac{\omega}{T_h}\Big).
\label{rev18}
\end{eqnarray}
 Comparing (\ref{rev18}), (\ref{rev16}) the temperature of BTZ black hole is given by,
\begin{eqnarray}
T_h=T\Big(1+\sum_i\beta_i\frac{\hbar^i}{r^i_+}\Big)^{-1},
\label{rev19}
\end{eqnarray}
where,
\begin{eqnarray}
T=(\frac{4}{\hbar}{\textrm{Im}}\int_C\frac{dr}{\sqrt {g_{tt}(r)g^{rr}(r)}})^{-1}= \frac{\hbar}{4}(Im \int_C{\frac{dr}{N^2(r)}})^{-1} 
\label{rev19a}
\end{eqnarray}
is the semiclassical Hawking temperature and other terms are corrections coming from the quantum effects.

 With this expression of modified Hawking temperature ($T_h$) the modified form of (\ref{1.09}) is,
\begin{equation}
dS_{{\textrm {bh}}}= \frac{dM}{T_h}-\frac{\Omega}{T_h}dJ.
\label{1.21}
\end{equation}
Here also $dS_{{\textrm {bh}}}$ is a perfect differential since the following relationship,
\begin{eqnarray}
\frac{\partial}{\partial J}\sum_i\frac{1}{T}\left(1+\frac{\beta_i\hbar^i}{r^i_+}\right)=\frac{\partial}{\partial M}\sum_i\frac{-\Omega}{T}\left(1+\frac{\beta_i\hbar^i}{r^i_+}\right)
\label{1.22}
\end{eqnarray}
 holds. This can be proved order by order by expanding the summation and substituting $r_+$, $\frac{1}{T}$ and $\frac{\Omega}{T}$ from (\ref{1.04}), (\ref{1.10}) and (\ref{1.11}) respectively.

We follow the same procedure as we did for the semiclassical case, to find the entropy in presence of quantum corrections. Here we use a new dictionary. The functional form of the dictionary does not change from (\ref{new6}) and we replace all the semiclassical quantities by their corrected forms, where necessary,
\begin{eqnarray}
x\rightarrow M,y\rightarrow J,P\rightarrow \frac{1}{T_h},Q\rightarrow \frac{-\Omega}{T_h},f\rightarrow S_{\textrm {bh}}.
\label{1.221}
\end{eqnarray}
Following this dictionary together with (\ref{1.14}) and using (\ref{rev19}) we obtain the following solution,
$$S_{\textrm {bh}}(M,J)= \int\frac{1}{T}\sum_i\left(1+\frac{\beta_i \hbar^i}{r^i_+}\right)dM + \int{\sum_i\frac{-\Omega}{T}\left(1+\frac{\beta_i \hbar^i}{r^i_+}\right)}dJ -$$
\begin{eqnarray}\int\frac{\partial}{\partial J}\left(\int\frac{1}{T}\sum_i\left(1+\frac{\beta_i \hbar^i}{r^i_+}\right)dM\right)dJ,
\label{1.23}
\end{eqnarray}
which is the modified version of (\ref{1.16}). It is possible to solve (\ref{1.23}) analytically for all orders . Let us now restrict upto second order corrections. In that case we expand the summation over the integrands in (\ref{1.23}) upto the second order, which yields,
$$S_{{\textrm {bh}}}(M,J)= \int\frac{1}{T}\left(1+\frac{\beta_1 \hbar}{r_+}+\frac{\beta_2 \hbar^2}{r^2_+}+{\cal O}(\hbar^3)\right)dM + \int\frac{-\Omega}{T}\left(1+\frac{\beta_1 \hbar}{r_+}+\frac{\beta_2 \hbar^2}{r^2_+}+{\cal O}(\hbar^3)\right)dJ -$$
\begin{eqnarray}\int\frac{\partial}{\partial J}\left(\int\frac{1}{T}\left(1+\frac{\beta_2 \hbar}{r_+}+\frac{\beta_2 \hbar^2}{r^2_+}+{\cal O}(\hbar^3)\right)dM\right)dJ.
\label{1.24}
\end{eqnarray}
Let us first solve the integral over $dM$ in (\ref{1.24}) by using the value of $r_+$ from (\ref{1.04}) which yields,
$$\int\frac{1}{T}\left(1+\frac{\beta_1\hbar}{r_+}+\frac{\beta_2\hbar^2}{r^2_+}+{\cal O}(\hbar^3)\right)dM= \frac{4\pi l}{\sqrt 2\hbar}(M+\sqrt{M^2-\frac{J^2}{l^2}})^{1/2}+ 2\pi\beta_1\log(M+\sqrt{M^2-\frac{J^2}{l^2}})-$$
\begin{equation}
\frac{4\sqrt2\pi\beta_2\hbar}{l^2(M+\sqrt{M^2-\frac{J^2}{l^2}})^{1/2}}+{\cal O}(\hbar^3).
\label{1.25}
\end{equation}
Using this result one can establish the following relationship between the integrands of the second and third integral of (\ref{1.24}), 
\begin{eqnarray}
\frac{\partial}{\partial J}\left(\int\frac{1}{T}\left(1+\frac{\beta_1 \hbar}{r_+}+\frac{\beta_2 \hbar^2}{r^2_+}+{\cal O}(\hbar^3)\right)\right)dM={\frac{-\Omega}{T}\left(1+\frac{\beta_1 \hbar}{r_+}+\frac{\beta_2 \hbar^2}{r^2_+}+{\cal O}(\hbar^3)\right)}.
\label{1.26}
\end{eqnarray}
Substituting this in (\ref{1.24}) and using (\ref{1.25}) we find the entropy of the BTZ black hole including quantum corrections,
$$ S_{{\textrm {bh}}}=\frac{4\pi l}{\sqrt 2\hbar}(M+\sqrt{M^2-\frac{J^2}{l^2}})^{1/2}+ 2\pi\beta_1\log(M+\sqrt{M^2-\frac{J^2}{l^2}})-$$
\begin{equation}
\frac{4\sqrt2\pi\beta_2\hbar}{l^2(M+\sqrt{M^2-\frac{J^2}{l^2}})^{1/2}}+{\cal O}(\hbar^3)+ const..
\label{1.27}
\end{equation}
Equation (\ref{1.27}) can be expressed in terms of horizon (outer) radius to yield,
\begin{equation}
S_{\textrm {bh}}= \frac{4\pi r_+}{\hbar}+4\pi\beta_1\log {r_+} -\frac{4\pi\beta_2\hbar}{l}(\frac{1}{r_+}) +{\cal O}(\hbar^3)+const..
\label{1.28}
\end{equation}
Substituting $r_+$ from (\ref{1.041}) and reinstating the unit $8G_3=1$, we can write this expression for entropy in terms of horizon area of the BTZ black hole as given by,
\begin{equation}
S_{\textrm {bh}}= \frac{A}{4\hbar G_3}+4\pi\beta_1\log A-\frac{64\pi^2\beta_2\hbar G_3}{l}(\frac{1}{A}) +{\cal O}(\hbar^3)+const..
\label{1.29}
\end{equation}
This can also be written in terms of semiclassical Bekenstein-Hawking entropy (\ref{1.1911}),
\begin{equation}
S_{\textrm {bh}}= S_{\textrm {BH}}+4\pi\beta_1\log(S_ {\textrm {BH}})-\frac{16\pi^2\beta_2}{l}(\frac{1}{{S_\textrm {BH}}}) +{\cal O}(\hbar^3)+const..
\label{1.30}
\end{equation} 
The first term in (\ref{1.30}) is the usual semiclassical Bekenstein-Hawking entropy (\ref{1.1911}) while the other terms are corrections due to quantum effects. We see that a logarithmic correction appears in the leading order as found before in \cite{Carlip, Govindarajan}.
 Incidentally, we also find inverse of area term in subleading order (\ref{1.29}) which was not discussed in other approaches \cite{Carlip,Govindarajan}.\\\\

{\it Discussions}:\\
The present analysis shows that the Hamilton-Jacobi variant of the tunneling method, which goes beyond the semiclassical approximations, developed in \cite{R.Banerjee2} for a spherically symmetric background, is equally applicable for other geometries like the BTZ example. The logarithmic correction to the area law was obtained in agreement with existing results \cite{Carlip, Govindarajan}. Next to leading corrections, not considered in \cite{Carlip,Govindarajan} were also found. Since the tunneling approach provides a direct intuitive picture of the Hawking effect, it is reassuring to observe that corrections to the semiclassical expressions are equally well describable here.

Note that there are two unknown parameters $\beta_1$ and $\beta_2$ appearing in (\ref{1.30}). We have still not been able to fix these parameters. However, following Hawking's original approach \cite{Hawk} based on the path integral regularization of zeta function, entropy gets a logarithmic correction to the semiclassical value and the coefficient of the logarithmic term is related with trace anomaly. A recent analysis \cite{Majhi} in the tunneling formalism based on the quantum WKB approach also reveals this fact. In \cite{Majhi} the parameter $\beta_1$ has been fixed for the Schwarzschild black hole under tunneling approach. To fix $\beta_1$ for BTZ black hole under tunneling approach  more generalised treatment is needed. However, in the present status  $\beta_2$ can not be fixed for any black hole.\\\\

{\it{Acknowledgment}}: The author thanks Rabin Banerjee for suggesting this investigation and Bibhas Ranjan Majhi for many useful discussions.  He also thanks the Council of Scientific and Industrial Research (CSIR), Government of India, for financial support.

\end{document}